\documentclass[twocolumn,showpacs,preprintnumbers,amsmath,amssymb,aps,nofootinbib]{revtex4}

\usepackage{graphicx}
\usepackage{amssymb}
\usepackage{epstopdf}
\usepackage{bm}

\begin{document}
    
\title{Using UHE Cosmic Rays to Probe the CBR and Test Standard Model Particle Physics}
 
\author{Frank~J.~Tipler and Daniel W. Piasecki}
\affiliation{Department of Mathematics, Tulane University, New Orleans, LA 70118}

\date{\today}

\begin{abstract}
Tipler has shown that if we assume that the particle physics Standard Model and DeWitt-Wheeler quantum gravity (equivalent to Feynman-Weinberg quantum gravity) are a Theory of Everything, then in the very early universe, the Cosmic Background Radiation (CBR) could not have coupled to right handed electrons and quarks.  Tipler further showed that if this property of CBR has continued, the Sunyaev-Zel'dovich (SZ) effect would be observed to be too low by a factor of two.  WMAP and PLANCK observed this.  Tipler showed that this CBR property would also mean the Ultra High Energy Cosmic Rays (UHECR) would propagate a factor of ten further than standard theory predicts, since most of the cross section for pion production when a UHECR hits a CBR photon is due to a quark spin flip, and such a flip cannot occur if a CBR particle cannot couple to right-handed quarks.  We show that taking this additional propagation distance into account allows us to identify the sources of 86\% of the UHECR seen by the Pierre Auger Collaboration.  We can also identify the sources of 9 of the 11 UHECR seen by the AGASA observatory, and the source of the 320 EeV UHECR seen by the Fly's Eye instrument.  We propose observations to test the theory underlying the UHECR identifications, beginning with measuring the redshifts of five galaxies whose apparent visual magnitude we estimate to be about 15, and whose positions we give to within one arcsecond.  The particle physics Standard Model identifies the Dark Energy and Dark Matter.

\end{abstract}

\maketitle

\section{Introduction}
Astrophysicists should analyze experimental data assuming the validity of the extensively tested laws of physics.  The Standard Model of particle physics is one such law; the Standard Model has been confirmed experimentally over the past 50 years (The Standard Model's most recent triumph has been explaining the muon g-factor; see \cite{Morishima 2018}).  According to the Standard Model, the electromagnetic field is not fundamental, but is instead composed of three more fundamental fields, an $SU(2)_L$ gauge field, a $U(1)$ gauge field, and a Higgs scalar field. Tipler has shown \cite{Tipler2005} that quantum field theory does not permit  $U(1)$ radiation (like the electromagnetic field) to exist in the very early universe.  Thus, initially, the CBR (Cosmic Background Radiation) must have been entirely $SU(2)_L$ radiation \cite{Tipler2005}. That is, initially, the CBR could only couple to left-handed fermions; this is the meaning of the subscript $L$.  If this property of the initial CMB has persisted to the present day, then the CBR would consist mainly of pseudo-photons, just like photons, but unable to couple to fermions of right-handed chirality.

This would mean that a CBR particle propagating through clusters of galaxies would interact with only half of the free electrons in the gas between the galaxies, so in 2005, Tipler predicted (\cite{Tipler2005}, pp. 945--950) that the SZ effect would be observed to be too low by a factor of two.  WMAP observed that the SZ effect was indeed too low (\cite{Lieuetal06}, \cite{DiegoPartridge10}), and PLANCK observed that it was too low by the predicted factor of two \cite{PLANCKXX13}.  \footnote{The PLANCK observers have since attempted to walk back their 2013 claim that the  SZ effect was too low by a factor of two, but we think they got it right the first time.}

A CBR composed mainly of pseudo-photons would mean that UHECR could propagate a factor of ten further through the CBR, since 90\% of the cross-section for pion production in the collision between a UHE proton and a CBR photon is due to a quark spin flip, and a pseudo-photon cannot generate a quark spin flip.  The GZK effect would still exist --- the pion production cross-section is still non-zero --- and it would still make its appearance at the energy predicted by Greisen \cite{Greisen} and by Zatsepin and Kuz'min \cite{ZK}.  However, the GZK effect would not be as strong, and this would explain why several UHECR observer groups initially claimed (e.g., \cite{Takeda2003}) the effect was absent.  It would also explain why UHECR with energies above the GZK cut-off have been seen by several groups.  UHECR would be able to propagate through the CBR composed mainly of pseudo-photons from a distance as great as $z=0.1$.  We shall use this fact to show that 86\% of the UHECR observed by the Pierre Auger Observatory can also be associated with Active Galactic Nuclei (AGN).  Standard theory can identity AGN sources for only 40\% of the UHECR observed by the Pierre Auger Observatory (\cite{Hand2010}, \cite{PierreAuger2010}). (Actually, standard CBR theory yields the identification of only 20\% if a $3^\circ$ window is used, the window size we propose.)  We predict that within $3^\circ$ of the arrival direction of the remaining unidentified UHECR, there will be found a previously unknown AGN. We in fact propose sources for an additional 6\% of the remaining unidentified UHECR, and show how these proposed sources can be confirmed by measuring the redshifts of galaxies whose positions we give to within one second of arc.  If these redshifts are all observed to be less than $z=0.1$, then we will have identified sources for 92\% of the Auger UHECR.

$\Lambda$CDM cosmology is often called ``Standard Model Cosmology,'' because this simple model is consistent with all observations. Tipler showed \cite{Tipler2005} that $\Lambda$ and Cold Dark Matter can naturally be explained by Standard Model particle physics.  When combined with standard quantum gravity (i.e, DeWitt-Wheeler quantum gravity, which is equivalent to Feynman-Weinberg quantum gravity if the appropriate boundary conditions are imposed), the particle physics Standard Model can explain (1) why the universe is observed to be classical rather than quantum, (2) why the universe is spatially flat, (3) why the universe is homogeneous and isotropic, (4) why the universe has more matter than antimatter, (5) what the Dark Matter is, and (6) what the Dark Energy is. 

The basic ideas of this paper will be presented in the next four pages.  The data will be presented in two sets of Tables, which will appear as appendices after the bibliography.

\section{Summary of Standard Model Cosmology}

Forty years of ever more precise observations have shown that the CBR has a Planckian distribution.  Most astrophysicists assume that a Planckian distribution means a field in thermal equilibrium, but this is false: a Planckian distribution need not arise from thermal process.  A Planckian distribution can also be a reflection of spacetime symmetries, as it is in the case of Hawking radiation, and in Rindler space, a spacetime that a spacetime with zero Riemann curvature everywhere that is globally the frame of a uniformly accelerating observer.  Tipler has has shown \cite{Tipler2005} that a quantized gauge field in a flat Friedmann universe necessarily has a Planckian distribution, with ``pseudo-temperature'' proportional to $1/R$, where $R$ is the scale factor of the Friedmann universe.  It is thus possible that the pure $SU(2)_L$ gauge field in the very early universe survived to the present (combined with the Higgs field), in which case the CBR would not electromagnetic thermal radiation, but instead would still be missing the $U(1)$ piece, which would mean that the CBR would not couple to right-handed quarks. The UHE cosmic ray observations over the past few decades, including the Auger observations, indicate that this is the case.

What has been holding up the advance of cosmology over the past thirty years is the assumption that observing a Planck spectrum must mean the radiation is in thermal equilibrium.  It is certainly true that if a massless boson radiation field is in thermal equilibrium, then the spectrum must be Planck.  Yet the converse need not be true.  Assuming the converse is true is a logical error called ``affirming the consequent.''  We cosmologists must learn to think ``outside the box'' (\cite{RootBernstein}, pp. 292--297) of identifying an observed Planck distribution with thermal equilibrium.

One might think that rejecting thermal equilibrium would mean rejecting the great triumphs of physical cosmology, the correct prediction of the helium and deuterium  abundances and the acoustic peak spectrum.  Not true.  These predictions follow not from thermal equilibrium, but from a consequence of thermal equilibrium, conservation of entropy. See for example any textbook of cosmology, e.g. \cite{Weinberg2008}, p. 150.  But the central point is this:  in  Tipler's theory, the CNR is PLANCKIAN, though at zero entropy, meaning that the Planck spectrum is imposed by the Friedman geometry rather than thermal equilibrium.  This in Tipler's theory, the particles created by the $SU(2)_L$ will be created by a non-thermal Planck distribution themselves, and this distribution will have the same pseudo-temprature as the $SU(2)_L$ field.  Adiabatic expansion  --- the standard cosmological assumption that yields the corrert observations --- will preserve these Planckian distributions.  Thus the only photon contribution to the CBR will be from the non-adiabatic process of re-ionization ending the Dark Ages.  About 10\% of the CBR would be photons today, and this is what is seen in SZ observations made by the PLANCK instrument.

Having a Planckian but non-thermal CBR in the very early universe allows the Standard Model to explain why there is more matter than antimatter in the universe.  The Standard Model has four fundamental fields, three gauge fields $U(1)$, $SU(2)_L$, and $SU(3)$ and a complex doublet scalar field, called the Higgs field.  The electromagnetic field is a combination of the first two gauge fields and the Higgs field, as is the weak force, while the $SU(3)$, the color force, is not coupled directly with the other three fields, and is responsible for the strong force.  Thus, there are a total of five fundamental forces: gravity and the four forces of the Standard Model.  A central theorem of quantum field theory, the Bekenstein Bound, requires isotropy and homogeneity at the Planck time, and also picks out the $SU(2)_L$ field as the only field that can exist at the Planck time.  Furthermore, this $SU(2)_L$ field must be self-dual, so it will force tunneling between the vacua of the electroweak force, generating only particles, and not anti-particles.  The Bekenstein Bound requires zero entropy and hence zero temperature at the Planck time.  This forces the Sakarov conditions to be obeyed in the early universe, even with the Planck distribution: the Planck spectrum comes from the Friedmann geometry, and not thermal equilibrium.

This quantum field mechanism of forcing the universe to be homogeneous eliminates primordial gravitational waves, which yields the prediction that B-modes in the CBR spectrum from this source will be absent.  This mechanism also shows why there must be super horizion over and under densities:  the Bekenstein Bound is a consistency condition that necessarily applies globally.  
 
Eddington \cite{Eddington1931}, then Lema\^itre \cite{Lemaitre1931} in 1931, and finally Feynman \cite{Feynman} in 1963, argued that the only natural initial condition for the universe was one of zero entropy and zero temperature.    A pure self dual $SU(2)_L$ at zero temperature field in the very early universe gives a new mechanism for Standard Model baryogenesis, and Tipler has shown \cite{Tipler2005} that this  mechanism naturally generates the observed photon to baryon ratio.  Furthermore, the created baryons and leptons (recall that $B - L$, baryon number minus lepton number, is conserved in SM baryogenesis) themselves break the homogeneity that the Bekenstein Bound requires at the Planck time, and in a spacetime that is very close to flat, generates the observed flat perturbation spectrum at the observed magnitude. 

Feynman and Weinberg showed many years ago that there is a unique renormalizable quantum theory of gravity for a spin two field, and we know gravity must be spin two from the Hulse-Taylor pulsar quadrupole energy loss rate.  Tipler has shown \cite{Tipler2005} that with the appropriate cosmological boundary conditions, the Feynman-Weinberg quantum theory of gravity is not only renormalizable, but term by term finite, and that the same mechanism that makes the theory term by term finite also forces the power series in the coupling constants to converge.  Applying this theory of quantum gravity to the pre-Planckian early universe implies that the wave function of the universe must have been a delta function at the initial singularity.  To see this, note that with appropriate boundary conditions, Feynman-Weinberg quantum gravity is equivalent to DeWitt-Wheeler quantum gravity \cite{HartleHawking1983}.  If the early universe is radiation dominated, then there is a choice of the matter conjugate variables which will order the three geometries by conformal time, and the DeWitt-Wheeler equation is then mathematically equivalent to the non-relativistic Schr\"odinger equation for the simple harmonic oscillator.  Requiring the initial wave function to be a delta function is equivalent to requiring that the Many-Worlds of quantum cosmology are classical from the initial singularity.  This solves the problem of ``why is the universe observed to be classical?'' 

The initial delta function wave function necessarily {\it explodes} outward, forcing the universe we find ourselves in to be flat, and the mechanism causing spatially flatness would thus be wave packet spreading, a process seen many times in elementary physics classes.  The universe's spatial flatness is due to a kinematical quantum mechanical mechanism, and not a finely tuned dynamical mechanism.  Tipler (\cite{Tipler1986a} p. 265, \cite{Tipler1986b} p. 212) proposed this quantum kinematical solution to the Flatness Problem in 1986, thrity-two years ago.   A kinematical mechanism is necessarily more stable than any dynamical mechanism.  An inflation field is not necessary to explain {\it any} cosmological observation.  Furthermore, this delta function mechanism of generating spatial flatness means that the perturbation spectrum will be scale invariant, because, as Harrison and Zel'dovich showed decades ago, only a scale invariant spectrum is allowed in a spatially flat universe.  Zero entropy means in addition that these scale invariant perturbations will be adiabatic.

The perturbation spectrum will actually not be perfectly scale invariant, because there are two ways of measuring scale, namely gravitational (the delta function forces this way to yield exact flatness) and Standard Model particle physics (this way gives a three-sphere closed universe which is quite large but not infinite).  The combination of these two ways imply that the perturbation spectrum will be merely ``nearly'' flat, so the scalar spectral index will be a bit less than one.

Thus the Standard Model of particle physics, when combined with standard quantum mechanics explains the observed isotropy, homogeneity, and spatial flatness of the universe, as well as the observed photon to baryon ratio, the observed excess of matter over anti-matter, and the Harrision-Zel'dovich perturbation spectrum.  And finally, it explains why we see the CZM  cut-off, and yet UHECRs with energies beyond the CZM cut-off.

Standard Model physics is not required to solve the Hubble Constant Problem, which is the $3.8\sigma$ difference \cite{Riess2018} between the Hubble constant measured by the distance ladder method --- $H_0 = 73.24 \pm 1.7 \,{\rm km/s/Mpc}$ \cite{Riess2016} --- and the Hubble constant measured by PLANCK --- $H_0 = 66.93 \pm 0.62\, {\rm km/s/Mpc}$.  We note that the inverse ladder method agrees with PLANCK \cite{Aubourg2015}, and this strongly suggests a systematic error in the distance ladder method.  Tipler \cite{Tipler1999} pointed out two decades ago that the distance ladder must be extended out to $z=3$ in order to be sure that one is in the Hubble flow.  As is well-known, $H_0$ will be biased toward higher values if we happen to live in an underdense region.

Using Standard Model physics to solve all of the problems of cosmology will mean accepting that the universe began in a singularity.  This many physicists are reluctant to do, because they believe that the existence of a singularity in a theory means the theory is wrong.  Nonsense!  Liouville's Theorem (bounded entire functions are constants) tells us that non-constant analytic functions necessarily have a singularity {\it somewhere}, and Liouville's Theorem is a consequence of the Fundamental Theorem of Calculus.  Do we therefore reject calculus? Of course not.  Instead, we make use of the singularities of the functions of complex analysis to solve problems.  We propose that physicsts should do the same for the cosmological singularity.  Experience tells us that there are no singularities in the laboratory.  Experience tells us nothing about the existence of singularities on the edge of space-time, the location of the cosmological singularity.  Tipler has recently shown \cite{Tipler2014} that the reason nature is quantized is to ensure that singularities do not appear in the laboratory.  Tipler has also shown \cite{Tipler1994} that quantization cannot eliminate singularities if gravity is curvature.

Conversely, accepting that the universe began in a singularity requires the use of quantum gravity theory in the ultra-high energy energy regime near the singularity, where energies were far beyond the Planck energy.  As mentioned above, applying standard quantum gravity theory there yields the solution to two outstanding cosmological problems:  (1) why does classical physics hold at the macroscopic level, and (2) why is the universe spatially flat?  

If the CBR is indeed composed of pseudo-photons and not photons, then the Standard Model of particle physics also tells us what the Dark Matter and Dark Energy are.  Tipler has pointed out that if the CBR were not thermalized in the early universe, then the Higgs field oscillations around the Higgs vacuum would be damped only by the expansion of the universe, and this oscillation energy has been known for decades (\cite{Tipler2005}. p.942) to have an effective energy density that falls off as ${\rm R}^{-3}$, which is to say, it would be the Dark Matter.  The Higgs vacuum energy minimum acts as an enormous negative cosmological constant, and this necessitates a positive cosmological constant to nearly cancel it out.  The sum of this positive cosmological constant and the Higgs vacuum energy is the Dark Energy.  Thus Standard Model particle physics and standard quantum gravity provide an explanation for all astronomical observations.  If the CBR is composed of a pseudo-photon field, it would represent a detectable fifth force, a fifth force whose existence arises from the more basic four forces of the Standard Model.

\section{Analysis of Pierre Auger and 
AGASA UHECR Data}

In the complete Pierre Auger data, there were 231 UHECR.  We used the online VizieR data base to scan the locations of these events in the 2006  (12th edition) of the Veron + list of Quasars and Active Galactic Nuclei, looking for quasars and AGN with a redshift up to $z=0.1$, as allowed by the possibility that the CMR would not couple to right-handed protons, and hence could propagate to Earth from this distance.  Following the original Pierre Auger procedure, we looked for acceptable sources within $3^\circ$ of the 231 UHECR directions. We are able to identify a source for 199, or $86\%$.  Conventional theory, with the window expanded to $5^\circ$,  allows identification of only $40\%$ (with the $3^\circ$ window, we have only 40, or $20\%$, with a redshift less than 0.01).  In addition, there is also a source ({\rm MCG} $+08.11.011$) located $3^\circ$ from the observed direction for the 320 EeV Fly's Eye event, a source noticed in 1994 by Elbert and Sommers \cite{Sommers}, who rejected this source since its redshift is $0.020$, a factor of 2 too high to be the source by conventional GZK theory, but well within the limit allowed by the Standard Model theory.  Of the identified UHECR sources, 107 of the 199, or $54\%$, are Seyfert galaxies.  Of the Seyferts, 45, or $42\%$, are Seyfert type 1, and thus 62, or $58\%$, are Seyfert type 2.  Most of the Seyferts, in other words, have broad doppler emission lines, which is what one would expect if the UHECR were emitted perpendicular to a supermassive black hole accretion disk.  

\section{Small Telescope Observations That Can Test the Theory}

Of the 32 UHECR for which we cannot identify a source, 13 are within $7^\circ$ of the galactic equator.  Because of the dust and gas in the plane of the galaxy, we would expect there would be fewer measured redshifts of galaxies lying near the galactic plane.  Therefore, we propose that within $3^\circ$ of the location of these 13 unidentified UHECR sources, there will be found an AGN with a redshift less than 0.1.  The locations to be searched are given in Table \ref{tab:UHECRunidentifiedlowgalactic latitude}.

Of the remaining 19 UHECR for which we cannot identify a source, there are 14 AGN in the WISE survey which are within $3^\circ$ of the direction of the UHECR all with reported redshift of 0.100, which means that their redshifts have been measured photometrically to within $0.1$, not that all of the galaxies have exactly the same redshifts to three decimal places \cite{MILLIQUAS17}.  We propose that astronomers measure the redshifts of these 14 WISE galaxies more precisely, and we predict that they all will have a redshift of less than 0.1.  The locations of these WISE galaxies are given to within an arcsecond in Table \ref{tab:UHECRunidentifiedhighgalactic latitudeWISEgalaxies}, together with our estimate of their apparent visual magnitude. 

Of the 14 WISE AGN, there are 5 which can be easily seen by observers in the northern hemisphere  (whom we assume have a pointing limit of $-30^\circ$).  Of these 5 northern AGN, one is in Auriga, one is in Capricornus, two are in Sagittarius, and one is in Canis Major.  Assuming that the 5 AGN have approximately the same magnitude in the visible that they have in the infrared, these 5 AGN all should have apparent visual magnitude 15.

The most easily observable of the 5 is the last entry in Table \ref{tab:UHECRunidentifiedhighgalactic latitudeWISEgalaxies}, our proposal for the source of an AGASA UHECR (observed April 9, 2002).  This WISE AGN is located, as indicated in the Table, at RA $05^h\, 46^m\,6^s$ and dec $+29^\circ\, 51^\prime\,46^{\prime\prime}$, in the constellation Auriga, about $4^\circ$ to the southwest of open cluster M37. 

The second most observable WISE AGN is Table \ref{tab:UHECRunidentifiedhighgalactic latitudeWISEgalaxies} entry day 030 of year 2009, located at RA $20^h\, 16^m\,08^s$ and dec $-17^\circ\, 00^\prime\,45^{\prime\prime}$, in the constellation Capricornus. about $2.5^\circ$ southwest of $\beta$ Capricorni.\footnote{Le Verrier predicted the existence of the planet Neptune in 1846 by assuming the validity of Newtonian theory in regions where his contemporaries believed it did not hold.  Le Verrier asked astronomers to look for Neptune ``about $5^\circ$ to the east of $\delta$ Capricorni'' (\cite{LeVerrier1846}, \cite{Grosser1962}, p. 111).  We are following Le Verrier, first by assuming the validity of confirmed physical law, and second, by deducing therefrom the existence of an interesting object in Capricorn.  The observations we propose cannot lead to the discovery of a new planet, but perhaps they can lead to the discovery of a new CBR: one composed of pseudo-photons rather than photons; to the discovery of the fifth force, a more important discovery than the discovery of a new planet.}

There are two of the 5 in the constellation Sagittarius.  One is Table \ref{tab:UHECRunidentifiedhighgalactic latitudeWISEgalaxies} entry day 191 of year 2009, located at RA $19^h\, 36^m\,08^s$ and dec $-20^\circ\, 43^\prime\,03^{\prime\prime}$.  The other is Table \ref{tab:UHECRunidentifiedhighgalactic latitudeWISEgalaxies} entry day 224 of year 2010, located at RA $18^h\, 53^m\,05^s$ and dec $-27^\circ\, 09^\prime\,34^{\prime\prime}$, located in the handle of the Teapot, less than $1^\circ$ southwest from $\sigma$ Sagittarii ($16^\circ$ from Sgr A*, so our galaxy's black hole is not the source of this UHECR).  Finally there is Table \ref{tab:UHECRunidentifiedhighgalactic latitudeWISEgalaxies} entry day 235 of year 2007, located at RA $07^h\, 00^m\,25^s$ and dec $-22^\circ\, 05^\prime\,29^{\prime\prime}$, in the constellation Canis Major, about $6^\circ$ southeast of Sirius.

\section{Discussion}

When the GZK effect was first discovered,  one UHECR had been seen with an energy above 100 EeV, and Greisen himself wrote \cite{Greisen} in his paper on the GZK effect that it was surprising such a cosmic ray could exist at all.  Seeing these few events beyond the GZK cut-off is evidence that the CMBR indeed cannot couple to right-handed protons.  Such a CMBR would imply the GZK cut-off --- which has now been seen --- but would also allow the existence of truly ultra high energy cosmic rays --- which have also been seen.

If the CBR is an $SU(2)_L$ gauge field combined with the Higgs vacuum, and not a complete electromagnetic field, then it cannot couple to right-handed electrons either.  Thus we would expect CBR pseudo-photons to show substantially less Sunyaev-Zel-dovich effect that conventional theory would predict.  This has now been seen, first by WMAP \cite{Lieuetal06}, \cite{DiegoPartridge10}, and later by PLANCK \cite{PLANCKXX13}.

Since these two completely different types of observations both indicate that the CBR may be composed  of pseudo-photons rather than photons, we suggest that the particle physics Standard Model coupled to DeWitt-Wheeler quantum gravity should be taken seriously as a Theory of Everything.  

As pointed out above, accepting the CBR as composed mainly of pseudo-photons rather than photons solve the two outstanding mysteries of cosmology, namely (1) what is the Dark Matter, and (2) what is the Dark Energy?  Both of these energies arise naturally from the particle physics Standard Model.

In addition. pseudo-photons would demonstrate that the Standard Model baryogenesis mechanism generated normal matter, and shows exaclty why there is more matter than anti-matter.  DeWitt-Wheeler quantum gravity forces the universe to be flat without having to involve any physics, like an inflaton field, not seen in the laboratory.  Standard quantum field theory as coded in the Bekenstein Bound, forces the universe to be homogeneous and isotropic, once again without having to appeal to physics not seen in the laboratory.

We think it is time to take standard physics, the physics confirmed in the laboratory in thousands of experiments, seriously.
\medskip
\section{Acknowledgments}

We thank P.~Sommers and Alan A. Watson for some very helpful comments on the Pierre Auger experiment, and for directing us to the published data on the arXiv.  


\clearpage

\appendix

\section{Tables of Unidentified UHECR from the Pierre Auger Data}

We first give tables that list the UHECR for which we have been unable to identify AGN sources.

\bigskip\bigskip\bigskip\bigskip\bigskip\bigskip\bigskip\bigskip

\pagebreak

\begin{table}
\caption{\label{tab:UHECRunidentified20042008} List of unidentified Pierre Auger UHECR for the years 2004 through 2008.  The first column gives the year of observation of the UHECR. The second column gives the day of observation.  The third gives the energy of the UHECR in EeV.   The fourth column gives the unidentified UHECR's galactic latitude. The fifth column gives the UHECR's galactic longitude.  Notice that of the 10 unidentified UHECR, 6 are within $5^\circ$ of the galactic equator.  It is hard to measure the redshift of an AGN near the galactic equator because of obscuration. }
\begin{ruledtabular}
\begin{tabular}{lcccc}
{\rm year} & {\rm day} &{\rm E} & {\rm galactic longitude} & {\rm galactic latitude}  \\

2004	 & 239 & $54.0$& $-59.1$ &$-31.8$ \\ 	 

2005	 & 233 &$61.9$ & $29.7$ &$+3.4$ \\ 

2006	 & 126 & $82.0$ & $57.6$ &$-4.7$ \\

2007	 & 205  & $61.9$ & $-55.9$ &$-0.6$ \\  

2007 & 235 & $60.8$ &$ -125.2 $& $ -7.7 $\\

2008	 & 013  & $64.2$ & $-1.9$ &$+13.7$ \\  

2008	 & 036  & $65.3$ & $-59.5$ &$-0.7$ \\ 

2008	 & 268  & $118.3$ & $36.5$ &$-3.6$ \\  

2008	 & 322  & $62.2$ & $-67.1$ &$-54.8$ \\ 

2008	 & 337  &  $65.8$ & $16.7$ &$+0.1$ \\ 

\end{tabular}
\end{ruledtabular}
\end{table}

\begin{table}
\caption{\label{tab:UHECRunidentified20092010} List of unidentified Pierre Auger UHECR for the years 2009 through 2010.  The first column gives the year of observation of the UHECR. The second column gives the day of observation.  The third gives the energy of the UHECR in EeV.   The fourth column gives the unidentified UHECR's galactic latitude. The fifth column gives the UHECR's galactic longitude.  Notice that of the 15 unidentified UHECR, only 4 are within $7^\circ$ the galactic equator.}
\begin{ruledtabular}
\begin{tabular}{lcccc}
{\rm year} & {\rm day} &{\rm E} & {\rm galactic longitude} & {\rm galactic latitude}  \\

2009	 & 030  & $66.2$ & $26.8$ &$-25.8$ \\ 

2009	 & 035  &  $57.7$ & $-54.2$ &$-23.1$ \\

2009	 & 191  & $59.5$  & $19.1$ &$-19.2$ \\

2009	 & 282  & $60.8$  & $168.6$ &$ -38.7$ \\

2009	 & 288  & $58.6$  & $+41.6$ &$8.4$ \\ 

2010 & 052 & $52.1$ & $-17.0$ &$-3.3$ \\

2010	 & 148  & $74.8$  & $-142.2$ &$ -17.5$ \\

2010	 & 196  & $52.3$  & $-32.6$ &$ -32.8$ \\

2010 & 224 & $65.2$ & $8.1$ &$-13.9$ \\   

2010 & 226 & $53.8$ & $71.2$ &$-25.0$ \\ 

2010 & 277 & $73.7$ & $-55.3$ &$-76.5$ \\ 

2010 & 320 & $54.3$ & $-86.2$ &$-34.1$ \\

2010 & 320 & $68.7$ & $-111.9$ &$0.4$ \\  

2010 & 347 & $54.9$ & $-36.7$ &$0.0$ \\

2010 & 348 & $54.4$ & $-61.9$ &$-6.2$ \\  

\end{tabular}
\end{ruledtabular}
\end{table}

\begin{table}
\caption{\label{tab:UHECRunidentified20122013} List of unidentified Pierre Auger UHECR for the years 2012 and 2013.  The first column gives the year of observation of the UHECR. The second column gives the day of observation.  The third gives the energy of the UHECR in EeV.   The fourth column gives the unidentified UHECR's galactic latitude. The fifth column gives the UHECR's galactic longitude.  Notice that of the 7 unidentified UHECR, 3 are within $3^\circ$ of the galactic equator.  It is hard to measure the redshift of an AGN near the galactic equator because of obscuration.}
\begin{ruledtabular}
\begin{tabular}{lcccc}
{\rm year} & {\rm day} &{\rm E} & {\rm galactic longitude} & {\rm galactic latitude}  \\

2012	 & 052  & $66.1$ & $-75.3$ &$-55.2$ \\ 

2012	 & 154  & $58.7$ & $-64.3$ &$-39.9$ \\ 

2012	 & 183  & $61.8$ & $-6.2$ &$2.7$ \\

2013	 & 036  & $73.6$ & $-34.8$ &$-19.7$ \\ 

2013	 & 222  & $61.5$ & $-41.3$ &$-12.1$ \\ 

2013	 & 332  & $65.2$ & $-30.5$ &$-1.0$ \\ 

2013	 & 364  & $53.2$ & $-54.5$ &$-1.2$ \\ 

\end{tabular}
\end{ruledtabular}
\end{table}

\begin{table}
\caption{\label{tab:UHECRunidentifiedhighgalactic latitude} List of all unidentified Pierre Auger UHECR with high galactic latitude (galactic latitudes which are NOT within $7^\circ$ of the galactic equator).  There are 19 such unidentified UHECR.  The first column gives the year of observation of the UHECR. The second column gives the day of observation.  The third gives the energy of the UHECR in EeV.   The fourth column gives the unidentified UHECR's galactic latitude. The fifth column gives the UHECR's galactic longitude.   }
\begin{ruledtabular}
\begin{tabular}{lcccc}
{\rm year} & {\rm day} &{\rm E} & {\rm galactic longitude} & {\rm galactic latitude}  \\

2004	 & 239 & $54.0$& $-59.1$ &$-31.8$ \\ 	 

2007 & 235 & $60.8$ &$ -125.2 $& $ -7.7 $\\

2008	 & 013  & $64.2$ & $-1.9$ &$+13.7$ \\  

2008	 & 322  & $62.2$ & $-67.1$ &$-54.8$ \\ 

2009	 & 030  & $66.2$ & $26.8$ &$-25.8$ \\ 

2009	 & 035  &  $57.7$ & $-54.2$ &$-23.1$ \\

2009	 & 191  & $59.5$  & $19.1$ &$-19.2$ \\

2009	 & 282  & $60.8$  & $168.6$ &$ -38.7$ \\

2009	 & 288  & $58.6$  & $+41.6$ &$8.4$ \\ 

2010	 & 148  & $74.8$  & $-142.2$ &$ -17.5$ \\

2010	 & 196  & $52.3$  & $-32.6$ &$ -32.8$ \\

2010 & 224 & $65.2$ & $8.1$ &$-13.9$ \\   

2010 & 226 & $53.8$ & $71.2$ &$-25.0$ \\ 

2010 & 277 & $73.7$ & $-55.3$ &$-76.5$ \\ 

2010 & 320 & $54.3$ & $-86.2$ &$-34.1$ \\

2012	 & 052  & $66.1$ & $-75.3$ &$-55.2$ \\ 

2012	 & 154  & $58.7$ & $-64.3$ &$-39.9$ \\ 

2013	 & 036  & $73.6$ & $-34.8$ &$-19.7$ \\ 

2013	 & 222  & $61.5$ & $-41.3$ &$-12.1$ \\

\end{tabular}
\end{ruledtabular}
\end{table}

\begin{table}
\caption{\label{tab:UHECRunidentifiedlowgalactic latitude} List of all unidentified Pierre Auger UHECR with LOW galactic latitude (galactic latitudes which are within $7^\circ$ of the galactic equator).  There are 13 such unidentified UHECR.  The first column gives the year of observation of the UHECR. The second column gives the day of observation.  The third column gives the UHECR galactic latitude. The fourth gives the UHECR's right ascension (J2000). The fifth column gives the UHECR's declination (J2000).  We predict an ADN with redshift less than $0.1$ will be found within $3^\circ$ of the positions in the Table.}
\begin{ruledtabular}
\begin{tabular}{lcccc}
{\rm year} & {\rm day} & {\rm GLAT} & {\rm right ascension} & {\rm declination}  \\

2005	 & 233 &$+3.4$ &$ 278.4$& $ -1.3$ \\ 	 

2006	 & 126 &$-4.7$ & $299.0$ &$19.4$ \\  

2007	 & 205 &$-0.6$ & $195.5$ &$-63.4$ \\ 

2008	 & 036 &$-0.7$ & $187.5$ &$-63.5$ \\

2008	 & 337 &$+0.1$  & $275.2$ &$-14.4$ \\

2009	 & 268 & $-3.6$   & $287.7$ &$1.5$ \\ 

2010	 & 052  & $-3.3$ & $258.1$ &$ -44.9$ \\

2010	 & 320 &$+0.4$  & $121.1$ &$ -30.6$ \\

2010 & 347& $+0.0$ & $231.9$ &$-56.6$ \\   

2010 & 348 & $-6.2$ & $179.7$ &$-61.9$ \\ 

2012 & 183 &$+2.7$ &$ 259.8$& $ -32.7$\\

2013	 & 332 & $-1.0$  & $241.6$ &$-53.5$ \\ 

2013	 & 364 & $-1.2$  & $198.8$ &$ -63.9$ \\

\end{tabular}
\end{ruledtabular}
\end{table}

\begin{table}
\caption{\label{tab:UHECRunidentifiedhighgalactic latitudeWISEgalaxies} List of all unidentified Pierre Auger UHECR with high galactic latitude AND which have a WISE galaxy within $3^\circ$ all which have a listed redshift of $0.100$. This means, as we pointed out in the text, that the redshift has been measured photometrically only to within $0.1$. The first column gives the year of observation of the UHECR. The second column gives the day of observation.  The third gives the WISEA number.  The fourth gives our estimate of the visual apparent magnitude of the WISEA galaxy.  Of the 19 UHECR with high galactic latitude, 14 have a possible WISEA source, leaving only 5 high galactic latitude UHECR with no candidate sources.   For those unidentified UHECR with no possible WISE source the galactic latitudes are given instead.  The J2000 right ascensions and declinations of the WISE objects are given in their names.   For example, the first object in the list, WISEA J070024.74-220528.8, has RA $07^{\rm h}$, $00^{\rm m}$, $24.74^{\rm s}$ and DEC $-22^\circ$, $05^\prime$, $28.8^{\prime\prime}$.  According to the WISE website,  these locations are not completely accurate, but nevertheless using the designation coordinates will result in an (unsigned) error of at most $0.0996^{\prime\prime}$ in declination, and $0.1488^{\rm s}$ in right ascension.  Accuracy to within a faction of a second should be sufficient to locate the AGN.  The last WISEA in the list is our proposed source for the last AGASA UHECR. }
\begin{ruledtabular}
\begin{tabular}{lccc}
{\rm year} & {\rm day} &{\rm WISEA number} & $\rm m_V$ \\

2004	 & 239 & $-31.8$ \\ 	 

2007 & 235 & WISEA J070024.74-220528.8 & $15^{HJK}$($20^{IR}$)\\

2008	 & 013  & $+13.7$ \\  

2008	 & 322  & WISEA J013930.20-624702.4 &  $12^{W}$($15^{IR}$)\\ 

2009	 & 030  & WISEA J201607.90-170044.5 & $15^{HJK}$($20^{IR}$)\\ 

2009	 & 035  &  WISEA J162218.92-843110.7 & $12^{W}$($17^{IR}$)\\ 

2009	 & 191  & WISEA J193608.18-204302.8 & $12^{W}$($17^{IR}$)\\

2009	 & 282  & $-38.7$ \\

2009	 & 288  & $+8.4$ \\ 

2010	 & 148  & $-17.5$ \\

2010	 & 196  & WISEA J201219.71-693603.7 & $15^{HJK}$($20^{IR}$) \\

2010 & 224 & WISEA J185304.60-270934.0 & $16^{HJK}$($21^{IR}$)\\   

2010 & 226 & SDSS J213738.10+182926.0 & 22 \\ 

2010 & 277 & WISEA J003708.21-414957.1 & $15^{HJK}$($20^{IR}$)\\

2010 & 320 & WISEA J053008.59-643617.6  & $16^{HJK}$($21^{IR}$)\\

2012	 & 052  & WISEA J020533.87-614820.7 & $13^{W}$($18^{IR}$) \\ 

2012	 & 154  & WISEA J024849.75-763942.5 & $15^{HJK}$($20^{IR}$) \\ 

2013	 & 036  & WISEA J174226.14-675411.0 & $17^{HJK}$($22^{IR}$) \\ 

2013	 & 222  & WISEA J155611.19-690351.7 & $15^{HJK}$($20^{IR}$) \\

2002 & 04 09 &WISEA J054606.42+295145.6 & $15^{HJK}$($20^{IR}$)\\

\end{tabular}
\end{ruledtabular}
\end{table}

\clearpage

\section{Tables of Identified UHECR from the Pierre Auger Data}

We now give lists for all the 199 UNECR for which we have been able to identify AGN as sources.
\bigskip

\pagebreak

\begin{table}
\caption{\label{tab:UHECR2004} List of Pierre Auger UHECR for 2004.  The first column gives the day of observation of the UHECR. The second gives the energy of the UHECR in EeV.  The third column is the most likely source, in our judgment.  The fourth column gives the astronomical classification of our proposed source. The fifth column gives the redshift of our proposed source.  The sixth column gives the angular distance between our proposed sources and the direction of the UHECR.  A question mark means we have not been able to identify a source within $3^\circ$ of the UHECR observed direction.}
\begin{ruledtabular}
\begin{tabular}{lccccc}
{\rm day} & E &{\rm source} & {\rm object type} & {\rm redshift} & {\rm distance} \\
\hline
125	 & $62.2$ &{\rm J17418-1212} & {\rm microquasar} &$0.037$ & $1.85^\circ$\\

142   & $84.7$  &$\rm J1304-3406$ & {\rm jet mode AGN} & $0.051$ &$2.96^\circ$\\

177 	& $54.6$      &\rm ESO 113-G10& {\rm Seyfert type 1.8} &$0.026$ &$2.65^\circ$\\

239	&$54.0$      &?&? &? &?\\

282	&$58.6$      &$\rm Centaurus\,B$&{\rm Seyfert type 1} &$0.0129$ &$0.796^\circ$\\

339	&$78.2$     & $\rm ESO 139-G12$&{\rm Seyfert type 2} &$0.017$&$2.21^\circ$\\

343	&$58.2$     &$\rm IC 4518A$ &{\rm Seyfert type 2}  &$0.016$ &$0.909^\circ$ \\
\end{tabular}
\end{ruledtabular}
\end{table}

\begin{table}
\caption{\label{tab:UHECR2005} List of Pierre Auger UHECR for 2005.  The first column gives the day of observation of the UHECR. The second gives the energy of the UHECR in EeV.  The third column is the most likely source, in our judgment.  The fourth column gives the astronomical classification of our proposed source. The fifth column gives the redshift of our proposed source.  The sixth column gives the angular distance between our proposed sources and the direction of the UHECR.  A question mark means we have not been able to identify a source within $3^\circ$ of the UHECR observed direction.}
\begin{ruledtabular}
\begin{tabular}{lccccc}
{\rm day} & E &{\rm source} & {\rm object type} & {\rm redshift} & {\rm distance} \\
\hline
050	 & $60.2$ &{\rm KUG 0202-122} & {\rm radio \,AGN} &$0.072$ & $2.88^\circ$\\

054   & $71.2$  &$\rm Tololo0109.383$ & {\rm Seyfert 1} & $0.0112$ &$0.371^\circ$\\

063 	& $71.9$      &\rm J22044-0056& {\rm AGN} &$0.063$ &$0.303^\circ$\\

081	&$51.1$      &\rm NGC 4945 &{\rm Seyfert 2} &$0.002$ &$2.02^\circ$\\

186	&$108.2$      &$\rm NGC 1194$&{\rm AGN} &$0.013$ &$0.673^\circ$\\

233	&$61.9$     & ?&? &?&?\\

295	&$54.9$     & ESO\, 344-G16 &{\rm Seyfert 1.5}  &$0.039$ &$0.852^\circ$ \\

306	&$74.9$     &\rm J07282339 &{\rm AGN}  &$0.0817$ &$2.44^\circ$ \\

347	&$77.5$     &NGC 452 &{\rm AGN}  &$0.017$ &$1.92^\circ$ \\
\end{tabular}
\end{ruledtabular}
\end{table}

\begin{table}
\caption{\label{tab:UHECR2006} List of Pierre Auger UHECR for 2006.  The first column gives the day of observation of the UHECR. The second gives the energy of the UHECR in EeV.  The third column is the most likely source, in our judgment.  The fourth column gives the astronomical classification of our proposed source. The fifth column gives the redshift of our proposed source.  The sixth column gives the angular distance between our proposed sources and the direction of the UHECR.  We have used the IAU truncation abbreviation convention for long object names.   A question mark means we have not been able to identify a source within $3^\circ$ of the UHECR observed direction.}
\begin{ruledtabular}
\begin{tabular}{lcccccc}
{\rm day} & E &{\rm source} & {\rm object type} & {\rm redshift} & {\rm distance} \\
\hline
005	 & $78.2$ &NPM1G-03.0065& AGN &$0.054$& $1.81^\circ$\\

035	 & $72.2$ &LQAC 053-007-015 & quasar &$0.097$& $0.845^\circ$\\

055	 & $52.8$ &ESO139-G12 & Seyfert 2 &$0.017$& $1.71^\circ$\\

064	 & $64.8$ & IC 4709 & Seyfert 2 &$ 0.0169 $& $1.008^\circ$\\

081	 & $69.5$ & WKK 2031 & Seyfert 2 &$0.031$ & $1.375^\circ$\\

100	 & $54.7$ & NPM1G-15.0089 & linear AGN &$0.0514$& $2.20^\circ$\\

118	 & $56.3$ &NGC 7069 & linear AGN &$0.031$&$0.504^\circ$\\

126	 & $82.0$ &? & ? &?& ?\\

142	 & $64.3$ & ESO 209-G12 & Seyfert1 &$0.040$& $2.78^\circ$\\
 
160	 & $60.7$ & IRAS 03278-4329 & Seyfert 2 & $0.058$& $0.217^\circ$\\

185	 & $89.0$ & NGC 7674  &  Seyfert 2 &$0.029$& $2.1^\circ$\\

263	 & $53.0$ & PGC 1439494 & Seyfert1 &$0.024$& $1.38^\circ$\\

284	 & $54.0$ &1Jy 0915-118 & linear AGN &$0.0547$& $2.92^\circ$\\

296	 & $67.7$ &MKN 607 & Seyfert 2 &$0.0092$& $2.26^\circ$\\

299	 & $59.5$ &NGC 5128  & Seyfert 2 &$0.001$& $2.32^\circ$\\

350	 & $60.0$ &NGC 6890 & Seyfert 2 &$0.008$&  $1.64^\circ$\\
\end{tabular}
\end{ruledtabular}
\end{table}

\newpage
\begin{table}
\caption{\label{tab:UHECR2007} List of Pierre Auger UHECR for 2007.  The first column gives the day of observation of the UHECR. The second gives the energy of the UHECR in EeV.  The third column is the most likely source, in our judgment.  The fourth column gives the astronomical classification of our proposed source. The fifth column gives the redshift of our proposed source.  The sixth column gives the angular distance between our proposed sources and the direction of the UHECR.  We have used the IAU truncation abbreviation convention for long object names.   A question mark means we have not been able to identify a source within $3^\circ$ of the UHECR observed direction.}
\begin{ruledtabular}
\begin{tabular}{lccccc}
{\rm day} & E &{\rm source} & {\rm object type} & {\rm redshift} & {\rm distance} \\
\hline
009	 & $53.8$ & UGC 11763 & Seyfert 1 &$0.061$& $2.89^\circ$\\

013	 & $127.1$ & 2MASX J1258 & AGN &$0.0473$& $1.71^\circ$\\

014	 & $52.2$ &Q 1241+1624 & Seyfert 1-2 &$0.023$& $1.99^\circ$\\

069	 & $60.0$ & NGC 5128 & Seyfert 2 &$ 0.001 $& $0.89^\circ$\\

084	 & $60.8$ & NGC 2907 & linear  AGN &$ 0.007 $& $1.42^\circ$\\

106	 & $70.3$ & MKN 975 & Seyfert 1 &$ 0.049 $& $1.00^\circ$\\

145	 & $68.4$ & NGC 1204 & Seyfert 2 &$ 0.015 $& $1.44^\circ$\\

161	 & $53.6$ & MARK 703 & AGN &$ 0.013 $& $2.67^\circ$\\

166	 & $54.9$ & PGC 3084749 & Seyfert 1 &$ 0.038 $& $2.82^\circ$\\
 
186	 & $61.5$ & IGR J14515-5542 & Seyfert 2 &$ 0.018 $& $2.69^\circ$\\

193	 & $79.7$ & NGC 7135 & AGN &$ 0.007 $& $2.14^\circ$\\

203	 & $57.0$ & PGC 3096554 & Seyfert 1 &$ 0.03 $& $2.18^\circ$\\

205	 & $61.9$ & ? & ? &$ ? $& $?^\circ$\\

221	 & $67.8$ & MKN 1376 & Seyfert 2 &$ 0.0061 $& $0.589^\circ$\\

227	 & $60.7$ & MCG -06.28.025 & linear AGN &$ 0.009 $& $1.58^\circ$\\

234	& $68.1$ & ESO 505-IG031 & Seyfert 2 &$0.04 $& $1.92^\circ$\\

235	& $60.8$ & ? & ? &$ ? $& $?^\circ$\\

295	& $65.9$ & LQAC-328-01$\ldots$ & AGN &$ 0.034 $& $2.86^\circ$\\

295	& $55.8$ & NGC 918 & AGN &$ 0.005 $& $2.86^\circ$\\

314	& $52.5$ & 2MASX J0353$\ldots$ & Seyfert 2 &$ 0.018 $& $1.42^\circ$\\

339	& $54.0$ & NGC 6240 & Seyfert 2 &$ 0.024 $& $2.96^\circ$\\

343	& $82.4$ & SDSS J052$\ldots$ & AGN &$0.100$& $1.16^\circ$\\

345	& $72.7$ & IRAS 205$\ldots$ & Seyfert 2 &$ 0.0239 $& $1.10^\circ$\\
\end{tabular}
\end{ruledtabular}
\end{table}

\begin{table}
\caption{\label{tab:UHECR2008} List of Pierre Auger UHECR for 2008.  The first column gives the day of observation of the UHECR. The second gives the energy of the UHECR in EeV.  The third column is the most likely source, in our judgment.  The fourth column gives the astronomical classification of our proposed source. The fifth column gives the redshift of our proposed source.  The sixth column gives the angular distance between our proposed sources and the direction of the UHECR.  (NELG means ``Narrow Emission Line Galaxy).  We have used the IAU truncation abbreviation convention for long object names.   A question mark means we have not been able to identify a source within $3^\circ$ of the UHECR observed direction.}
\begin{ruledtabular}
\begin{tabular}{lccccc}
{\rm day} & E &{\rm source} & {\rm object type} & {\rm redshift} & {\rm distance} \\
\hline
010	 & $80.2$ & NGC 6500 & linear AGN &$$0.01 & $2.05^\circ$\\

013	 & $64.2$ & ? & ? &$?$& $?^\circ$\\

018	 & $111.8$ &NPM1G-19.0685 & Seyfert 2 &$0.031$& $2.05^\circ$\\

036	 & $65.3$ & ? & ? &$ ? $& $?^\circ$\\

048	 & $60.4$ & 2QZ J01$\ldots$.1-27$\ldots$ & NELG &$ 0.053 $& $2.37^\circ$\\

049	 & $56.0$ & NGC 1566 & Seyfert 1 &$ 0.0043$& $2.26^\circ$\\

051	 & $53.3$ & IRAS 13120-5453 & Seyfert 2 &$ 0.031 $& $1.85^\circ$\\

052	 & $56.2$ & MCG -02-15-004 & AGN &$ 0.029 $& $2.60^\circ$\\

072	 & $52.4$ & IRAS 12031-3216 & Seyfert 2 &$ 0.039 $& $2.62^\circ$\\
 
087	 & $73.1$ & NGC 5643 & Seyfert 2 &$ 0.0033 $& $2.23^\circ$\\

118	 & $62.9$ & 2MASXi J0716$\ldots$ & QSO &$ 0.052 $& $1.70^\circ$\\

142	 & $56.7$ & MKN 1347 & Seyfert 1 &$ 0.050 $& $2.06^\circ$\\

184	 & $55.7$ & PGC 1425207 & Seyfert 2 &$ 0.042 $& $2.31^\circ$\\

192	 & $55.1$ & IC 4995	 & Seyfert 2 &$ 0.016 $& $2.63^\circ$\\

205	 & $56.7$ & MCG +02.60.017 & Seyfert 2 &$ 0.026 $& $1.98^\circ$\\

250	& $52.0$ & LQAC 068+002 & AGN &$ 0.016 $& $1.59^\circ$\\

264	& $89.5$ & ESO 208-G34 & AGN &$ 0.025 $& $1.08^\circ$\\

266	& $61.2$ & 6dFGS g22345 $\ldots$ & AGN &$ 0.056 $& $2.45^\circ$\\

268	& $118.3$ & ? & ? &$ ? $& $?^\circ$\\

282	& $58.1$ & MCG -03.34.064 & Seyfert 2 &$ 0.017 $& $1.63^\circ$\\

296	& $64.7$ & ESO 541-G001  & Seyfert 2 &$ 0.021 $& $2.65^\circ$\\

322	& $62.2$ & ? & ? &$ ? $& $?^\circ$\\

328	& $63.1$ & MCG 1-22-013 & AGN &$ 0.047 $& $2.87^\circ$\\

329	& $66.9$ & MCG -01.05.047 & Seyfert 2 &$ 0.017 $& $0.98^\circ$\\

331	& $52.6$ & IRAS 20033-2803 & AGN &$ 0.047 $& $2.98^\circ$\\

337	& $65.8$ & ? & ? &$ ? $& $?^\circ$\\

355	& $71.1$ & IGR J13168-7157 & Seyfert 1 &$ 0.07 $& $2.40^\circ$\\

362	& $74.0$ & NGC 5357 & linear AGN &$ 0.016 $& $1.11^\circ$\\
\end{tabular}
\end{ruledtabular}
\end{table}

\begin{table}
\caption{\label{tab:UHECR2009} List of Pierre Auger UHECR for 2009.  The first column gives the day of observation of the UHECR. The second gives the energy of the UHECR in EeV.  The third column is the most likely source, in our judgment.  The fourth column gives the astronomical classification of our proposed source. The fifth column gives the redshift of our proposed source.  The sixth column gives the angular distance between our proposed sources and the direction of the UHECR. We have used the IAU truncation abbreviation convention for long object names.   A question mark means we have not been able to identify a source within $3^\circ$ of the UHECR observed direction.}
\begin{ruledtabular}
\begin{tabular}{lccccc}
{\rm day} & E &{\rm source} & {\rm object type} & {\rm redshift} & {\rm distance} \\
\hline
007	 & $61.0$ & PGC 3082731 & AGN &$0.058$& $1.62^\circ$\\

030	 & $66.2$ & ? & ? &$?$& $?^\circ$\\

032	 & $70.3$ &MCG -03.01.002 & AGN &$0.036$& $1.42^\circ$\\

035	 & $57.7$ & ? & ? &$ ? $& $?^\circ$\\

039	 & $64.1$ & NGC 2989 & AGN &$ 0.013 $& $0.81^\circ$\\

047	 & $52.9$ & NPM1G-18.0222 & AGN &$ 0.042 $& $2.47^\circ$\\

051	 & $66.7$ & NGC 5253 & AGN &$ 0.001 $& $1.91^\circ$\\

073	 & $72.5$ & MCG -06.28.025 & linear AGN &$ 0.009 $& $0.236^\circ$\\

078	 & $74.4$ & LQAC\_125-057 & galaxy &$ 0.060 $& $2.74^\circ$\\
 
078	 & $59.0$ & NGC 613 & AGN &$ 0.005 $& $2.78^\circ$\\

080	 & $65.8$ & 6dFGS g1635550$\ldots$ & AGN* &$ 0.0004 $& $2.03^\circ$\\

080	 & $63.8$ & IRAS 11215-2806 & Seyfert 2 &$ 0.014 $& $1.35^\circ$\\

083	 & $56.2$ & PGC 1399638 & Seyfert 1 &$ 0.047 $& $2.89^\circ$\\

140	 & $55.1$ & TEX 2149-084 & Seyfert 2 &$ 0.035 $& $2.73^\circ$\\

160	 & $52.8$ & LQAC\_045-023 & Seyfert 1 &$ 0.035 $& $2.41^\circ$\\

162	& $70.5$ & IC 1813 & AGN &$ 0.015 $& $1.45^\circ$\\

163	 & $71.9$ & NGC 625 & AGN &$ 0.001 $& $1.26^\circ$\\

172	 & $65.8$ & CGMW 4-1205 & Seyfert 2 &$ 0.065 $& $1.46^\circ$\\

191	& $59.5$ & ? & ? &$ ? $& $?^\circ$\\

197	 & $52.2$ & SDSS J084518$\ldots$ & Seyfert 1 &$ 0.061 $& $2.06^\circ$\\

202	 & $63.6$ & IC 1524 & Seyfert 1 &$ 0.019 $& $2.10^\circ$\\

212	& $55.3$ & ESO 18-G09 & Seyfert 2 &$ 0.017 $& $1.04^\circ$\\

219	 & $53.2$ & NGC 788 & Seyfert 2 &$ 0.014 $& $2.04^\circ$\\

219	 & $58.3$ & IRAS 20253-8152 & Seyfert 2 &$ 0.034 $& $0.66^\circ$\\

237	& $70.0$ & UGC 11805SW & two galaxies &$ 0.018 $& $0.53^\circ$\\

250	 & $52.3$ & UGC 9035 & linear AGN &$ 0.027 $& $0.69^\circ$\\

262	 & $58.7$ & RX J0319.8-2627 & Seyfert 1 &$ 0.076 $& $0.58^\circ$\\

274	& $82.3$ & IC 4870 & AGN &$ 0.003 $& $2.95^\circ$\\

281	 & $75.3$ & IRAS 17080+1347 & AGN* &$ 0.031 $& $0.89^\circ$\\

282	 & $60.8$ & ? & ? &$ ? $& $?^\circ$\\

288	& $58.6$ & ? & ? &$ ? $& $?^\circ$\\

304	 & $55.6$ & IRAS F11500-0211 & Seyfert 1 &$ 0.0035 $& $2.57^\circ$\\

335	 & $52.5$ & H 1118-429 & Seyfert 1 &$ 0.0567 $& $0.96^\circ$\\
\end{tabular}
\end{ruledtabular}
\end{table}

\begin{table}
\caption{\label{tab:UHECR2010} List of Pierre Auger UHECR for 2010.  The first column gives the day of observation of the UHECR. The second gives the energy of the UHECR in EeV.  The third column is the most likely source, in our judgment.  The fourth column gives the astronomical classification of our proposed source. The fifth column gives the redshift of our proposed source.  The sixth column gives the angular distance between our proposed sources and the direction of the UHECR. We have used the IAU truncation abbreviation convention for long object names.   A question mark means we have not been able to identify a source within $3^\circ$ of the UHECR observed direction.}
\begin{ruledtabular}
\begin{tabular}{lccccc}
{\rm day} & E &{\rm source} & {\rm object type} & {\rm redshift} & {\rm distance} \\
\hline
024	 & $54.3$ & IGR J06415+3251 & Seyfert 2 &$0.017$& $2.97\circ$\\

045	 & $61.5$ & HE 1136-2304 & Seyfert &$0.027$& $2.19^\circ$\\

050	 & $64.5$ &IRAS 15091-2107 & Seyfert 1 &$0.044$& $0.18^\circ$\\

052	 & $66.9$ & ? & ? &$ ? $& $?^\circ$\\

072	 & $72.9$ & PGC 1365707 & Seyfert 2 &$ 0.019 $& $1.88^\circ$\\

121	 & $54.7$ & 6dFGS g08093$\dots$ & AGN* &$ 0.0004 $& $2.29^\circ$\\

148	 & $74.8$ & ? & ? &$ ? $& $?^\circ$\\

182	 & $82.0$ & NGC 5084 & linear AGN &$ 0.005 $& $2.75^\circ$\\

193	 & $58.4$ & SDSS J1001$\ldots$ & AGN &$ 0.043 $& $2.17^\circ$\\
 
194	 & $53.8$ & PGC 1365707 & Seyfert 2 &$ 0.019 $& $2.77^\circ$\\

196	 & $52.3$ & ? & ? &$ ? $& $?^\circ$\\

204	 & $53.2$ & Q 1209-1105 & AGN &$ 0.016 $& $2.46^\circ$\\

205	 & $53.5$ & IRAS 20253-8152 & Seyfert 2 &$ 0.034 $& $1.08^\circ$\\

223	 & $56.1$ & PGC 259433 & AGN &$ 0.090 $& $2.60^\circ$\\

224	 & $65.2$ & ? & ? &$ ? $& $?^\circ$\\

226	& $75.6$ & ? & ? &$ ? $& $?^\circ$\\

235	 & $60.3$ & Circinus Galaxy & Seyfert 2 &$ 0.0014 $& $1.64^\circ$\\

238	 & $69.6$ & B1514-24 & QSO &$ 0.049 $& $2.95^\circ$\\

239	& $58.4$ & NPM1G-15.0552 & AGN &$ 0.080 $& $1.60^\circ$\\

256	 & $76.1$ & IRAS 08417-1351 & Seyfert 1 &$ 0.028 $& $1.69^\circ$\\

277	 & $73.7$ & ? & ?&$ ? $& $?^\circ$\\

284	& $89.1$ & IRAS 14167-7236 & Seyfert 1 &$ 0.026 $& $2.25^\circ$\\

295	 & $58.0$ & ESO112-6 & AGN* &$ 0.029 $& $2.46^\circ$\\

310	 & $53.1$ & PGC 1351981 & Seyfert 1 &$ 0.047 $& $2.74^\circ$\\

311	& $70.5$ & Carafe Nebula & Linear AGN &$ 0.016 $& $2.37^\circ$\\

319	 & $55.0$ & IGR J07597-3842 & Seyfert 1 &$ 0.040 $& $1.69^\circ$\\

320	 & $54.3$ & ? & ? &$ ? $& $?^\circ$\\

320	& $68.7$ & ? & ? &$ ? $& $?^\circ$\\

342	 & $54.6$ & ESO 265- G 023 & Seyfert 1 &$ 0.057$& $0.689^\circ$\\

347	 & $54.9$ & ? & ? &$ ? $& $?^\circ$\\

348	& $54.4$ & ? & ? &$ ? $& $?^\circ$\\

364	 & $68.0$ & NGC 3621 & AGN &$ 0.002 $& $2.75^\circ$\\
\end{tabular}
\end{ruledtabular}
\end{table}

\begin{table}
\caption{\label{tab:UHECR2011} List of Pierre Auger UHECR for 2011.  The first column gives the day of observation of the UHECR. The second gives the energy of the UHECR in EeV.  The third column is the most likely source, in our judgment.  The fourth column gives the astronomical classification of our proposed source. The fifth column gives the redshift of our proposed source.  The sixth column gives the angular distance between our proposed sources and the direction of the UHECR.  AGN* means that we cannot independently identify the object as an AGN, but we list the object as an AGN because it is in the VizieR data base of quasars and AGN.  We have used the IAU truncation abbreviation convention for long object names.}
\begin{ruledtabular}
\begin{tabular}{lccccc}
{\rm day} & E &{\rm source} & {\rm object type} & {\rm redshift} & {\rm distance} \\
\hline
019	 & $64.4$ & IGR J18027-1455 & Seyfert 1 &$0.035$& $2.24^\circ$\\

026	 & $100.1$ & IRAS 09595-0755	 & Seyfert 1 &$0.055$& $2.21^\circ$\\

035	 & $54.0$ &ESO 506-G04 & Linear AGN &$0.013$& $0.41^\circ$\\

038	 & $58.2$ & LQAC 033-02$\dots$ & Seyfert 1 &$ 0.051 $& $2.42^\circ$\\

041	 & $52.0$ & LQAC\_125$\ldots$ & AGN* &$ 0.060 $& $2.13^\circ$\\

045	 & $62.7$ & MCG -2-37-004 & Seyfert 2 &$ 0.041 $& $2.10^\circ$\\

049	 & $60.3$ & PGC 1237895 & Seyfert 1 &$ 0.033 $& $1.52^\circ$\\

075	 & $71.1$ & Q 1515+0205 & Quasar &$ 0.020 $& $0.93^\circ$\\

086	 & $56.2$ & MCG-01.27.031 & Seyfert 1 &$ 0.021 $& $2.41^\circ$\\
 
106	 & $81.4$ & WISEA J2027$\ldots$ & AGN* &$ 0.100 $& $2.13^\circ$\\

111	 & $69.7$ & MARK 585 & AGN &$ 0.021 $& $1.34^\circ$\\

113	 & $54.8$ & IGR J19405-3016 & Seyfert 1 &$ 0.052 $& $2.59^\circ$\\

119	 & $67.3$ & LQAC 255-0$\ldots$ & AGN* &$ 0.033 $& $2.83^\circ$\\

120	 & $73.1$ & PGC 1439494 & Seyfert 1 &$ 0.023 $& $1.82^\circ$\\

132	 & $56.8$ & NGC 1097 & Linear AGN &$0.004 $& $1.86^\circ$\\

136	& $65.9$ & IRAS 22547-8018 & Seyfert 2 &$ 0.038 $& $2.17^\circ$\\

162	 & $55.9$ & SDSS J0902$\ldots$ & AGN &$ 0.030 $& $2.93^\circ$\\

203	 & $77.9$ & LQAC\_125-057$\ldots$ & AGN* &$ 0.060 $& $2.54^\circ$\\

207	& $56.4$ & MCG-3-58-7 & Seyfert 2 &$ 0.031 $& $2.09^\circ$\\

215	 & $68.3$ & NVSS J16313 $\dots$ &  AGN* &$ 0.072 $& $2.38^\circ$\\

221	 & $70.8$ & NGC 2845 & Seyfert 2 &$ 0.008 $& $2.22^\circ$\\

240	& $58.8$ & NGC 5643 & Seyfert 2 &$ 0.003 $& $2.38^\circ$\\

252	 & $80.9$ & WISEA J1853$\ldots$ & AGN* &$ 0.100 $& $1.41^\circ$\\

294	 & $75.6$ & TOL 0514-415 & Seyfert 2 &$ 0.049 $& $1.42^\circ$\\

307	& $52.4$ & IC 1339 & Linear AGN &$ 0.028 $& $1.62^\circ$\\

309	 & $63.3$ & ESO353-G9 & Seyfert 2 &$ 0.016 $& $2.84^\circ$\\

316	 & $70.2$ & LQAC 002-035$\dots$ & Seyfert 1 &$ 0.093 $& $2.91^\circ$\\

318	& $57.2$ & NGC 2992 & Seyfert 2 &$ 0.008 $& $2.71^\circ$\\

360	 & $67.4$ & IAU 2031-359 & AGN* &$0.088$& $2.93^\circ$\\

361	 & $92.8$ & FRL 357 & AGN &$ 0.028 $& $2.47^\circ$\\

364	 & $64.8$ & NGC 5291 & AGN &$ 0.014 $& $1.28^\circ$\\
\end{tabular}
\end{ruledtabular}
\end{table}

\begin{table}
\caption{\label{tab:UHECR2012} List of Pierre Auger UHECR for 2012.  The first column gives the day of observation of the UHECR. The second gives the energy of the UHECR in EeV.  The third column is the most likely source, in our judgment.  The fourth column gives the astronomical classification of our proposed source. The fifth column gives the redshift of our proposed source.  The sixth column gives the angular distance between our proposed sources and the direction of the UHECR.   A question mark means we have not been able to identify a source within $3^\circ$ of the UHECR observed direction.}
\begin{ruledtabular}
\begin{tabular}{lccccc}
{\rm day} & E &{\rm source} & {\rm object type} & {\rm redshift} & {\rm distance} \\
\hline
012	 & $62.4$ & 3C 29.0 & AGN &$0.045$& $2.33^\circ$\\

052	 & $66.1$ & ? & ? &$?$& $?^\circ$\\

081	 & $99.0$ &5BZG J2103-6812 & AGN* &$0.041$& $2.84^\circ$\\

103	 & $70.4$ & PKS0959-443 & Quasar &$ 0.021 $& $2.94^\circ$\\

109	 & $62.6$ & IC 1816 & Seyfert 2 &$ 0.017 $& $2.78^\circ$\\

132	 & $58.5$ & MKN 1330 & Seyfert 1 &$ 0.009 $& $0.87^\circ$\\

154	 & $58.7$ & ? & AGN* &$ ? $& $?^\circ$\\

155	 & $60.0$ & LQAC 244-028\_001 & AGN* &$ 0.055 $& $2.49^\circ$\\

162	 & $83.8$ & ESO 543-G11 & Seyfert 1 &$ 0.086 $& $2.97^\circ$\\
 
183	 & $61.8$ & ? & ? &$ ? $& $?^\circ$\\

189	 & $61.1$ & ESO 244-17 & Seyfert 1 &$ 0.024 $& $1.90^\circ$\\

193	 & $54.4$ & NPM1G-04.0637 & Seyfert 1 &$ 0.025 $& $2.35^\circ$\\

206	 & $56.8$ & IRAS 20253-8152 & Seyfert 2 &$ 0.034$& $1.41^\circ$\\

211	 & $58.7$ & MCG 2-30-017 & AGN &$ 0.021$& $2.27^\circ$\\

301	 & $53.3$ & NGC 1410 & Seyfert 2 &$ 0.025 $& $2.24^\circ$\\

332	& $71.1$ & NGC 5852 & AGN &$ 0.022 $& $1.34^\circ$\\

\end{tabular}
\end{ruledtabular}
\end{table}

\begin{table}
\caption{\label{tab:UHECR2013} List of Pierre Auger UHECR for 2013.  The first column gives the day of observation of the UHECR. The second gives the energy of the UHECR in EeV.  The third column is the most likely source, in our judgment.  The fourth column gives the astronomical classification of our proposed source. The fifth column gives the redshift of our proposed source.  The sixth column gives the angular distance between our proposed sources and the direction of the UHECR.   A question mark means we have not been able to identify a source within $3^\circ$ of the UHECR observed direction.}
\begin{ruledtabular}
\begin{tabular}{lccccc}
{\rm day} & E &{\rm source} & {\rm object type} & {\rm redshift} & {\rm distance} \\
\hline
011	 & $55.7$ & ESO 511-G030 & Seyfert 1 &$0.022$& $2.95^\circ$\\

027	 & $62.7$ & MCG -06.30.015 & Seyfert 1 &$0.008$& $2.65^\circ$\\

027	 & $70.7$ &PKS 0352-686 & BL Lac &$0.087$& $0.96^\circ$\\

031	 & $53.2$ & 5BZG J2103-6812 & AGN* &$ 0.041 $& $1.03^\circ$\\

036	 & $73.6$ & ? & ? &$ ? $& $?^\circ$\\

052	 & $71.9$ & NGC 1692 & AGN &$ 0.035 $& $0.08^\circ$\\

070	 & $53.9$ & PGC 917316 & Seyfert 1 &$ 0.058 $& $2.42^\circ$\\

119	 & $62.1$ & NGC 2824 & AGN &$ 0.009 $& $1.02^\circ$\\

132	 & $57.3$ & ESO012-G21 & Seyfert 1 &$ 0.033 $& $2.87^\circ$\\
 
134	 & $85.3$ & MCG-01.22.006 & Seyfert 2 &$ 0.023 $& $2.70^\circ$\\

144	 & $54.3$ & MCG-07.05.010 & Linear AGN &$ 0.017 $& $0.64^\circ$\\

163	 & $52.2$ & NGC 7733 & Seyfert 2 &$ 0.034 $& $2.85^\circ$\\

175	 & $58.9$ & CGCG 74-129 & Seyfert 2 &$ 0.016 $& $2.05^\circ$\\

190	 & $68.8$ & ESO 055-IG02 & Seyfert 2 &$ 0.048 $& $2.91^\circ$\\

191	 & $67.3$ & ESO 340-22 & Seyfert 2 &$ 0.056 $& $2.03^\circ$\\

222	& $61.5$ & ? & ? &$ ? $& $?^\circ$\\

224	 & $63.4$ & MCG -02.58.021 & AGN &$ 0.024 $& $0.43^\circ$\\

247	 & $84.8$ & 6dFGS g1011$\ldots$ & AGN* &$ 0.100 $& $0.53^\circ$\\

249	& $55.5$ & NGC 3281 & Seyfert 2 &$ 0.011 $& $2.02^\circ$\\

249	 & $65.4$ & IRAS 06317-6403 & Seyfert 2 &$ 0.048 $& $2.61^\circ$\\

281	 & $58.5$ & IRAS 21363-2700 & Seyfert 1 &$ 0.030 $& $2.96^\circ$\\

297	& $73.0$ & IAU 1045-721 & AGN* &$ 0.026 $& $1.83^\circ$\\

302	 & $54.6$ & IRAS 19542+1110 & LIRG &$ 0.065 $& $2.54^\circ$\\

319	 & $54.4$ & 1RXS J190749$\ldots$ & Seyfert 1 &$ 0.073 $& $2.70^\circ$\\

320	& $52.9$ & 1RXS J185650$\dots$ & AGN &$ 0.056 $& $1.48^\circ$\\

329	 & $63.6$ & MCG-02.31.015 & Linear AGN &$ 0.018 $& $1.81^\circ$\\

332	 & $65.2$ & ? & ? &$ ? $& $?^\circ$\\

352	& $72.5$ & ESO 121-IG28 & Seyfert 2 &$ 0.040 $& $2.19^\circ$\\

364	 & $53.2$ & ? & ? &$ ? $& $?^\circ$\\

\end{tabular}
\end{ruledtabular}
\end{table}

\begin{table}
\caption{\label{tab:UHECR2014} List of Pierre Auger UHECR for 2014.  The first column gives the day of observation of the UHECR. The second gives the energy of the UHECR in EeV.  The third column is the most likely source, in our judgment.  The fourth column gives the astronomical classification of our proposed source. The fifth column gives the redshift of our proposed source.  The sixth column gives the angular distance between our proposed sources and the direction of the UHECR.}
\begin{ruledtabular}
\begin{tabular}{lccccc}
{\rm day} & E &{\rm source} & {\rm object type} & {\rm redshift} & {\rm distance} \\
\hline
008	 & $60.0$ & ESO 033-G02 & Seyfert 2 &$0.018$& $2.07^\circ$\\

030	 & $74.5$ & ESO 381-G07 & Seyfert 1 &$0.055$& $0.92^\circ$\\

032	 & $54.6$ &ESO 506-G04 & Linear AGN &$0.013$& $1.38^\circ$\\

049	 & $54.9$ & LQAC 001-050$\dots$ & Seyfert 1 &$ 0.033 $& $1.01^\circ$\\

059	 & $60.2$ & IGR J15415-5029 & AGN &$ 0.032 $& $2.94^\circ$\\

064	 & $63.8$ & MS 03215-6657 & Quasar &$ 0.093 $& $2.36^\circ$\\

065	 & $118.3$ & UGC 12237 & Seyfert 2 &$ 0.028 $& $2.89^\circ$\\

\end{tabular}
\end{ruledtabular}
\end{table}

\begin{table}
\caption{\label{tab:UHECRAGASA} List of all AGASA UHECR,  There were 11 detected from 1993 to 2002.  The first column gives the year of observation of the UHECR. The second column gives the day of observation.  The third gives the energy of the UHECR in EeV.   The third column is the most likely source, in our judgment.  The fourth column gives the astronomical classification of our proposed source. The fifth column gives the redshift of our proposed source.  The sixth column gives the angular distance between our proposed sources and the direction of the UHECR.   Of the 11 UHECR, we can identify the source of 9.  We do not include the WISEA object in the 9, since the redshift is not measured with sufficient precision.  We have used the IAU truncation abbreviation convention for long object names.   A question mark means we have not been able to identify a source within $3^\circ$ of the UHECR observed direction.}
\begin{ruledtabular}
\begin{tabular}{lcccccc}
{\rm year} & {\rm day} &{\rm E} & {\rm source} & {\rm type} & {\rm redshift} & {\rm distance}  \\

1993	 & 01 21 & 101 & SDSS J0820 & AGN &$0.044$ &$0.87^\circ$ \\ 	 

1993	 & 12 03 & 213 & PG 0119+229 & Seyfert 1 &$0.053$ &$2.69^\circ$ \\

1994	 & 07 06 & 134 & 3C 388 & Seyfert &$0.091$ &$2.70^\circ$ \\

1996	 & 01 11 & 144 & PGC 1740204 & AGN &$0.041$ &$2.61^\circ$ \\

1996	 & 10 22 & 105 & ? &$?$ &$?^\circ$ \\ 

1997	 & 03 30 & 150 & CGMW 3-4394 & Seyfert 1 &$0.010$ &$0.41^\circ$ \\

1998	 & 06 12 & 120 & NGC 7479 & Seyfert 2 &$0.008$ &$2.70^\circ$ \\
 
1999	 &09 22 & 104 & LQAC\_346 & AGN* &$0.019$ &$0.69^\circ$ \\\

2001	 & 04 30 & 122 & NGC 3941 & Seyfert 2 &$0.003$ &$1.93^\circ$ \\

2001	 & 05 10 & 246 & MRK 331 & Seyfert 2 &$0.018$ &$1.76^\circ$ \\
 
2002	 & 04 09 & 121 & WISEA J0546 & AGN* &$0.100$ &$2.37^\circ$ \\

\end{tabular}
\end{ruledtabular}
\end{table}






\end{document}